# EXPERIMENTAL STUDY FOR THE FEASIBILITY OF A CRYSTALLINE UNDULATOR

S. Bellucci[1], S. Bini[1], V.M. Biryukov[2], Yu.A. Chesnokov[2], S. Dabagov[1], G. Giannini[1], V. Guidi[3], Yu.M. Ivanov[4], V.I. Kotov[2], V.A. Maisheev[2], C. Malagù[5], G. Martinelli[5], A.A. Petrunin[4], V.V. Skorobogatov[4], M. Stefancich[5], D. Vincenzi[5]

[1]*INFN - Laboratori Nazionali di Frascati, P.O. Box 13, 00044 Frascati, Italy*
[2]*Institute for High Energy Physics, 142281 Protvino, Russia*
[3]*Department of Physics and INFN, Via Paradiso 12, I-44100 Ferrara, Italy*
[4]*St. Petersburg Institute for Nuclear Physics, Russia*
[5]*Department of Physics and INFM, Via Paradiso 12, I-44100 Ferrara, Italy*

**Abstract**

We present an idea for creation of a crystalline undulator and report its first realization. One face of a silicon crystal was given periodic micro-scratches (trenches) by means of a diamond blade. The X-ray tests of the crystal deformation due to given periodic pattern of surface scratches have shown that a sinusoidal shape is observed on both the scratched surface and the opposite (unscratched) face of the crystal, that is, a periodic sinusoidal deformation goes through the bulk of the crystal. This opens up the possibility for experiments with high-energy particles channeled in crystalline undulator, a novel compact source of radiation.

**Corresponding author:** V.M. Biryukov, Institute for High Energy Physics, Protvino, Moscow region, Ru-142281 Russia; FAX: +7 0967 742867; E-mail: biryukov@mx.ihep.su

The wavelength λ of a photon emitted in an undulator is in proportion to the undulator period $L$ and in inverse proportion to the square of the particle Lorentz factor γ. The minimal period $L$ achieved presently with the electromagnetic undulators is limited to several millimeters [1], with respective restriction on the photon energy in the order of $\hbar\omega=2\pi\hbar\gamma^2 c/L$. The crystalline undulators (CU) with periodically deformed crystallographic planes offer huge electromagnetic fields and could provide a quite short period $L$ of an undulator in sub-millimeter range. This way one can also arrange for substantial amplitudes $A$ of oscillation for the particles channeled through the undulator and thus increase the intensity of the radiation.

Currently, bent crystals are largely used for channeling extraction of 70-GeV protons at IHEP (Protvino) with efficiency reaching 85% at intensity well over $10^{12}$ particle per second, steered by silicon crystal of just 2 mm in length [2]. A bent crystal (5 mm long) is installed into the Yellow ring of the Relativistic Heavy Ion Collider where it channels *Au* ions and polarized protons of 100-250 GeV/u as a part of the collimation system [3].

With a strong world-wide attention to novel sources of radiation, there has been broad theoretical interest to compact crystalline undulators, with some approaches covering also nanotechnology to make use of nanotubes to guide radiating particles [4-12]. The properties of crystal undulator radiation are reviewed, e.g., in refs. [10-12].

Kaplin et al. first proposed the idea of crystalline undulator in 1979 [4,5]; the same authors have suggested the historically first idea [5] how one could realize a CU, namely they proposed to apply ultrasonic waves for periodical deformation of crystal lattice. More recent idea [11] was the usage of graded-composition strained layers in superlattices; in particular, in *$Ge_xSi_{1-x}$* superlattices the concentration of *Ge*, *x*, is varied periodically in order to achieve a periodical deformation of crystal lattice basing on the technique demonstrated in Ref. [13]. Another recently discussed idea [14] is to provoke periodical strains in the crystal by covering its surface by thin narrow periodically placed strips of $Ge_xSi_{1-x}$ grown onto a *Si* substrate. However these ideas are still pending realization.

In bent crystal channeling experiments at IHEP Protvino with 70-GeV protons, it was found that accidental micro-scratches on a crystal surface caused a deformation of the crystallographic planes to substantial depths, down to a few hundred microns as depicted in Fig. 1(a). The picture of the plane parallelism violation can be reconstructed through analysis of the profile data of 70-GeV protons channeled in crystals (ref. [15], p.120). This analysis shows that the protons in the vicinity of scratches are retained in channeling mode but do experience a substantial angular deviation following the deformation of the crystal planes. Therefore, this effect could be profitably used for creation of CU by making a periodic series of micro trenches on the crystal surface as shown on Fig. 1(b).

For the first experimental proof of the method, a special diamond blade scratched one face of a silicon plate by a set of parallel trenches (grooves). A sample with dimensions of 50 x 17 x 0.48 mm$^3$ was prepared from commercial silicon wafer. The large polished faces of the sample were parallel to crystallographic planes (0 0 1), other faces were parallel to planes (0 1 1) and (0 1 -1). On one of the large faces of the sample, 16 grooves were made with a period of 1 mm along the 50 mm edge. The grooves had 100 μm width and the same depth.

The prepared sample was carefully investigated with the help of the single-crystal diffractometer being part of the two-crystal X-ray spectrometer described in [16]. The set-up is shown in Fig. 2. The X-rays emitted (*Mo K$_{\alpha 1}$* X-rays with 17.4 keV energy were used) by the tube anode were collimated into a beam of 2 mm height and 40 μm width and incident onto the surface of a sample positioned on a table that could be translated with the accuracy of about 1 μm. The table was mounted on the rotation axis of a standard theodolit ensuring the angular accuracy of 1 arcsec. A NaI counter with wide acceptance window detected the diffracted radiation. The count rate of diffracted X-ray quanta is maximal under Bragg condition, achieved by rotation of the sample.

Two parameters were measured for each X-ray beam position: - intensity of diffracted beam; - angular position of the sample corresponding to a half intensity of the diffracted beam.

The intensity is proportional to the integral reflectivity of the crystal at the point of X-ray beam diffraction, and is very sensitive to crystal deformations. In present studies, the sensitivity of this parameter to the bending of the crystallographic planes in the area of diffraction was a fraction of arcsec.

The angular position of the sample is related through Bragg law to the orientation of the crystallographic planes and the magnitude of interplanar distance; the accuracy of angle readings was a fraction of arcsec. The change in the second parameter along x-axis is completely due to the change in the orientation of crystallographic planes. Sensitivity of this parameter to plane orientation was on the level of 2 arcsec in our measurements.

Fig.3 (a, b) shows the measured intensities and angles as functions of the beam incidence position at the surface with grooves. One can see that the angular deformations of the crystal planes reach an order of 50 μrad, and the corresponding amplitude of plane deformation is in the order of 80 Å; the deformation amplitude is obtained from integration of the angle-versus-position function of Fig. 3 (b).

Fig. 4 (a, b) shows the corresponding values for the opposite (unscratched) face of the crystal. One can see that on this side the local periodic deformations are reduced to 15 µrad with amplitudes on the order of 25 Å, and are smoothed in the shape (being closer to sinusoidal). In the central part of the crystal, the deformation amplitudes on the order of 50 Å were reached (if interpolated between the two faces).

As proven by the present experimental study, it is feasible to manufacture a crystalline undulator for channeling of radiating particles with a sub-millimeter period and with the amplitude of channel oscillations in the order of 50 Å, which is the range of parameters of most interest for a novel radiation source[10-12]. Based on this success, an experiment on photon emission in a crystalline undulator is being planned for the LNF Frascati.

**Acknowledgements**.

This work was partially supported by INFN - Gruppo V, as NANO experiment, by INTAS-CERN Grant No. 132-2000 and RFBR Grant No. 01-02-16229, by the ì Young Researcher Projectî of the University of Ferrara, and by the Russian Federation Federal Program "Integration".

# Figure captions

**Figure 1**

(a) Angular distortion of the crystal planes near a surface scratch (trench).

(b) Scheme of proposed crystalline undulator.

**Figure 2**

Scheme of one-crystal mode of reflection-type measurements.

**Figure 3**

(a) Intensity of diffracted beam versus the position of the incident X-ray beam along $x$-axis, for crystal side with grooves.

(b) Rotation angle of the sample versus the position of the incident X-ray beam along $x$-axis, for crystal side with grooves.

**Figure 4**

(a) Same as in Fig. 3(a), but for the crystal side without grooves.

(b) Same as in Fig. 3(b), but for the crystal side without grooves.

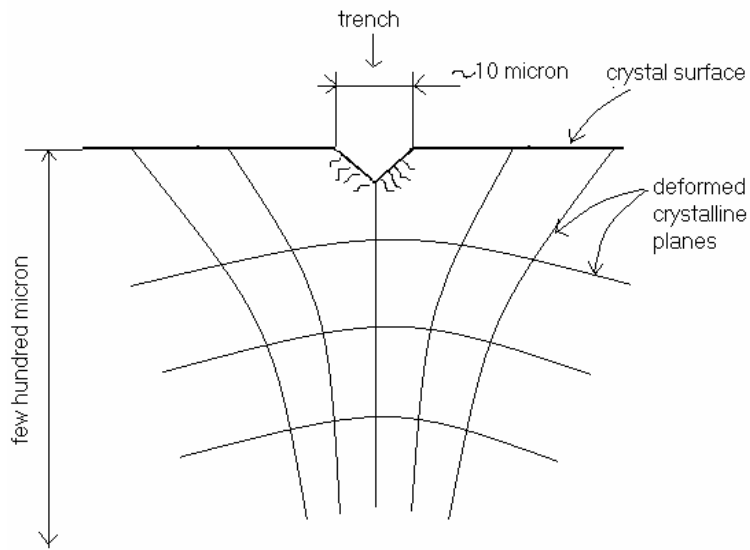

**Figure 1 (a)**

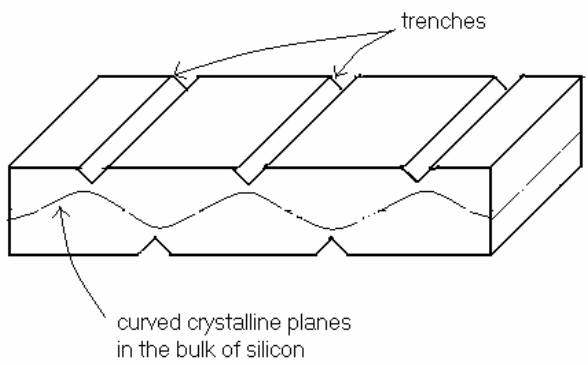

**Figure 1 (b)**

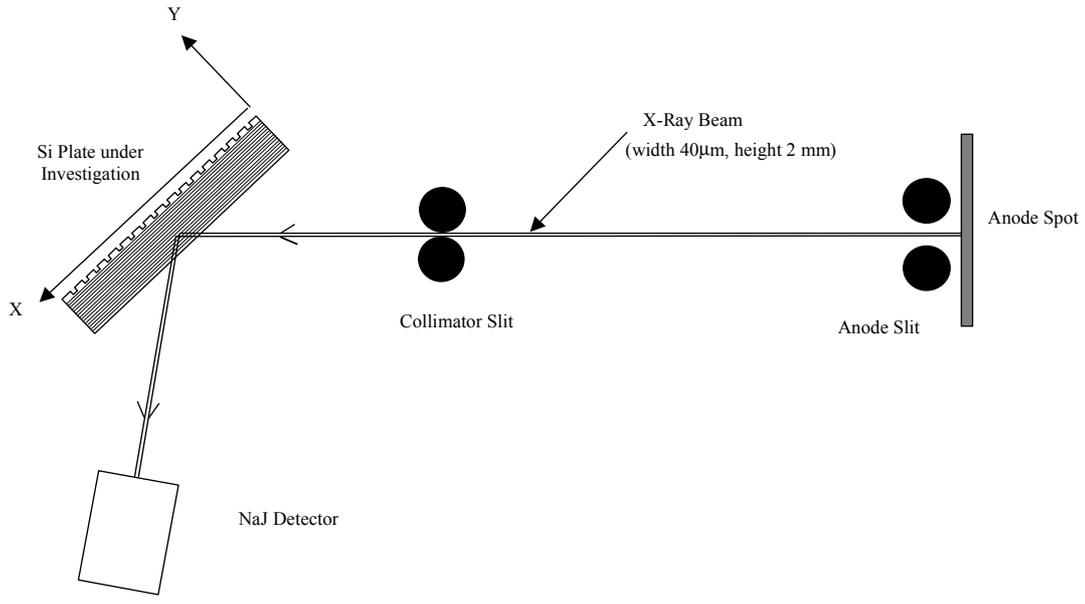

**Figure 2**

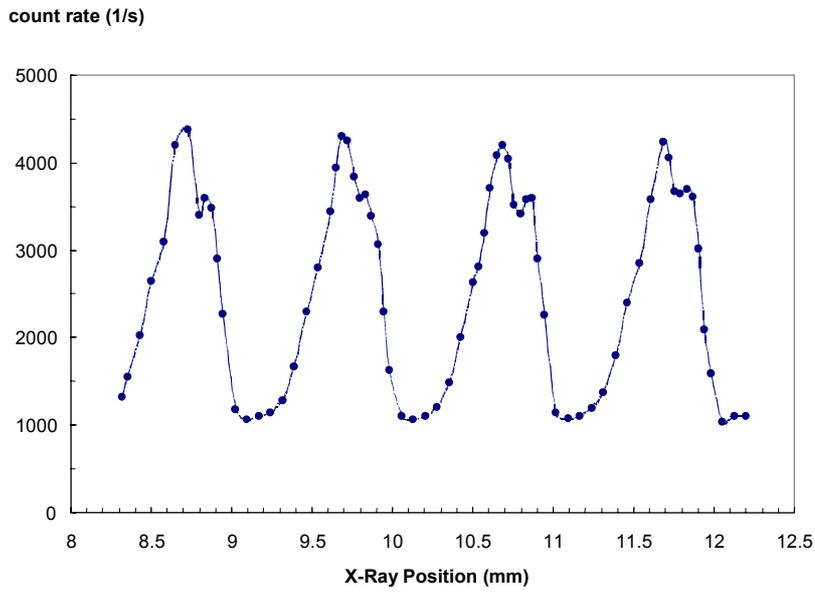

**Figure 3 (a)**

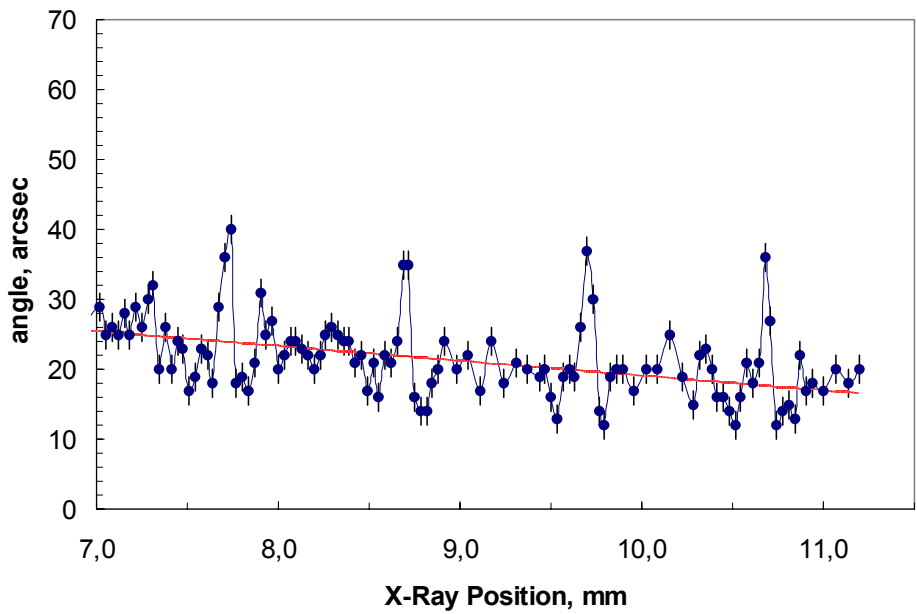

**Figure 3 (b)**

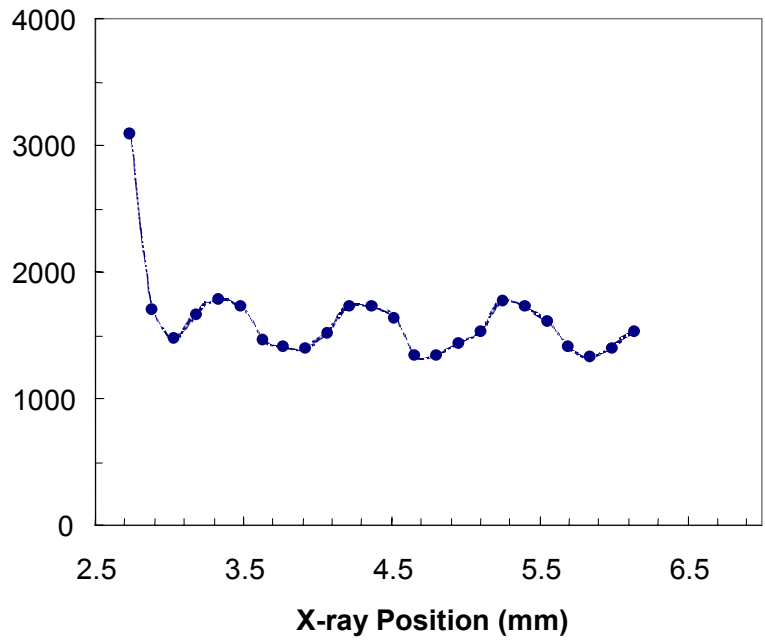

**Figure 4 (a)**

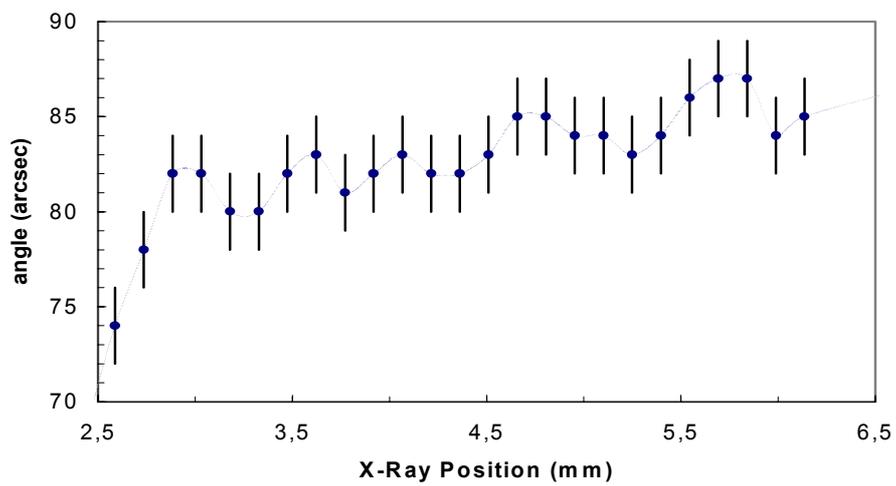

**Figure 4 (b)**